\documentclass[11pt,a4]{article}          

\usepackage[T1]{fontenc}

\usepackage{graphicx}
\usepackage{mathptmx}      
\usepackage{flushend}
\usepackage{wasysym}
\usepackage[numbers,sort&compress]{natbib}
\usepackage[colorlinks,citecolor=blue,urlcolor=blue,linkcolor=blue]{hyperref}

\usepackage{lineno}

\begin{document}
\title{Diffractive Bremsstrahlung at High-$\beta^\star$ LHC \\ {\large Case Study}
\author{
J. J. Chwastowski\footnote{Corresponding author, e-mail: Janusz.Chwastowski@ifj.edu.pl}\ , S. Czekierda,
R. Staszewski, M. Trzebi\'nski\\[8pt]
The Henryk Niewodnicza\'nski Institute of Nuclear Physics,\\
Radzikowskiego 152,
31-342 Krak\'ow, Poland
}
}

\maketitle

\begin{abstract}
Feasibility studies of the measurement of the exclusive diffractive bremsstrahlung cross-section in proton-proton scattering at 
the centre of mass energy of 13 TeV at the LHC are reported. Present studies were performed for the low luminosity LHC running 
with the betatron function value of 90~m using the ATLAS associated forward detectors ALFA and ZDC. A simplified approach 
to the event simulation and reconstruction is used. The background influence is also discussed. 
\end{abstract}

\section{Introduction}
\begin{figure}[h]
\begin{minipage}{\columnwidth}
\centering
\makebox[0.35\columnwidth][l]{a)}
\hspace{0.1\columnwidth}
\makebox[0.35\columnwidth][l]{b)}\\
\includegraphics[width=0.35\columnwidth]{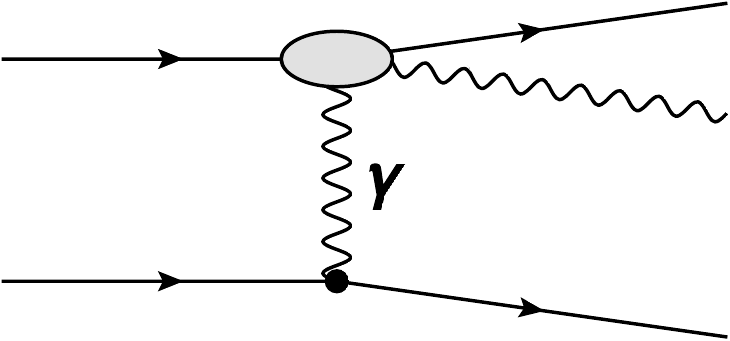}
\hspace{0.1\columnwidth}
\includegraphics[width=0.35\columnwidth]{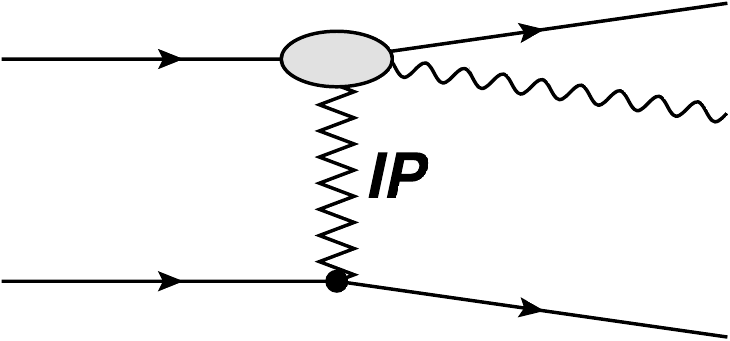}
\makebox[0.35\columnwidth][l]{\ }
\end{minipage}
\caption{The bremsstrahlung process diagrams: a) elecctromagnetic process, b) diffractive process. The blobs indicate a mixture of various mechanisms 
of photon emission.}
\label{fig:brem_graph}
\end{figure}

Electromagnetic bremsstrahlung off nuclei is commonly used in various applications. In the domain of high energy physics it was used to produce the 
tagged photon beam to study photoproduction (see for example \cite{photoprod}).  At HERA, the lepton induced  process, 
$e+p \rightarrow e+p+\gamma$, was  not only used to determine the machine absolute and instantaneous luminosities but also served as an efficient 
beam diagnostic and monitoring tool  (see  for example \cite{zeus}). The measurements exploited basic properties of the electromagnetic bremsstrahlung 
final state: an approximate energy conservation as well as the angular properties of radiated photons. Feynman diagram of this process is shown in 
Fig.~\ref{fig:brem_graph}a.

In 2011, V. Khoze and others \cite{khoze1} postulated to enrich the LHC physics programme with the measurement of the high energy photons 
radiated in the elastic proton-proton scattering  -- the study of exclusive diffractive bremsstrahlung  depicted in 
Fig.~\ref{fig:brem_graph}b. In particular, such a measurement can be considered as complementary to the absolute luminosity monitors proposed in 
\cite{khoze2, krasny1}. Both, electromagnetic and diffractive bremsstrahlung  processes are mediated by soft colourless exchanges with diffractive 
bremsstrahlung being driven by the exchange of a pomeron.  Also, the sum of the radiated photon and the scattered proton energies is nearly equal to 
the radiating particle energy due to the softness of the exchanged pomeron.

P. Lebiedowicz and A. Szczurek \cite{szczurek} extended the work of V. Khoze and others by introducing the proton form-factor into the calculations 
and by considering also other mechanisms leading to the $p\,p\,\gamma$ final state, for example the virtual photon re-scattering. They concluded that
these processes do not play a very important role and can be safely neglected at high energies. The values of the parameters in their model are subject to 
experimental verification. In particular, the model uses the Donnachie-Landshoff parameterisation \cite{dl} with linear pomeron trajectory 
$\alpha(t) = 1.0808 + 0.25\cdot t$. The elastic scattering slope evolution is  described by 
$B(s) = B^{NN}_{I\!P}+2\cdot\alpha^\prime_{I\!P}\ln{s/s_0}$ where $s_0 = 1$~GeV$^2$ and $B^{NN}_{I\!P} = 9$~GeV$^{-2}$. For a full account of 
the model details and its parameters a reader is referred to original publications. It is worth mentioning that in this model the absorptive are effectively 
taken into account and therefore there is no need to apply the gap survival probability factor.

In this paper a feasibility study of the diffractive photon bremsstrahlung measurement in proton-proton interactions at the presently available centre of 
mass energy of 13 TeV at the LHC is reported. It is assumed that the LHC runs with low instantaneous luminosity and the value of the betatron function of 
90 metres. The study focuses on the ATLAS  detector~\cite{atlas} and considers  the bremsstrahlung photon registrations in the ATLAS very forward Zero 
Degree Calorimeter (ZDC) \cite{zdc} and that of the scattered protons in the ALFA stations \cite{alfa} of ATLAS. 

It is also worth noting that this process has never been studied at high energies, however, feasibility studies performed for STAR experiment at RHIC 
energies \cite{app} and also the LHC \cite{app2} running with low value of the betatron function\footnote{The betatron function defines the distance 
measured from a given point along the orbit, after which the beam dimensions in the transverse plane to the motion are doubled.} at the interaction point, 
$\beta^\star = 0.55$ m, show a non-negligible potential of the measurement. In the latter case the use of  the ATLAS main detector and the ATLAS 
Forward Proton (AFP) \cite{afp} was considered. Both studies suggest that the measurements are possible if the accelerator runs with small or moderate 
instantaneous luminosity. Yet, another study was performed also at $\sqrt{s} = 10$ TeV and for the $\beta^\star = 90$~m \cite{khoze1} for the TOTEM 
where the authors postulated to tag the elastic events with bremsstrahlung photons. Such data can be used to measure the $p_T$ dependence of the 
elastic cross-section as well as the ratio of the elastic to total cross-section.

\section{Experimental Set-up}

\subsection{Accelerator}

Presently, the LHC delivers the proton beams accelerated to 6500~GeV and runs with high instantaneous luminosity. The operation mode of the machine 
can be described by a set of parameters defining the properties of its magnetic lattice -- the so-called accelerator optics. The nominal or ``collision'' optics 
was foreseen to be associated with $\beta^\star = 0.55$~m, however, presently the accelerator runs routinely with $\beta^\star$ value of 0.4~m. Below 
the so-called 90~m optics is considered. It should be stressed that such runs are dedicated to the study of elastic scattering at the LHC and are typically 
performed every year. During these data collection periods ATLAS includes the ALFA detector into the data acquisition chain. Table \ref{tab:param} 
presents the selected parameters of the proton beams at the ATLAS IP for the beam emittance\footnote{The beam emittance measures the spread of the 
beam particles in the momentum-position phase space.}, $\epsilon = 3.75$~$\mu$rad\,m, $\beta^\star = 90$~m optics and the beam energy 
$E_{beam} = 6500$~GeV.

\begin{table}[h]
\centering
\caption{The beam parameters at the ATLAS interaction point: the angular dispersion, the beam transverse size at the detector location and the crossing 
angle for the collision optics for the beam energy $E_{beam} = 6500$~GeV.} 
\label{tab:param}

\begin{tabular*}{\columnwidth}{@{\extracolsep{\fill}}ccc@{}}
 {Angular}  &	{Beam transverse}&Beam crossing\\
 {dispersion [$\mu$rad] } & {size [mm]}  &  half angle [$\mu$rad]\\
\hline
2.5& 0.21 & 0\\
\hline
\end{tabular*}
\end{table}

\begin{figure*}
\begin{minipage}{\textwidth}
\centering
\includegraphics[width=\textwidth]{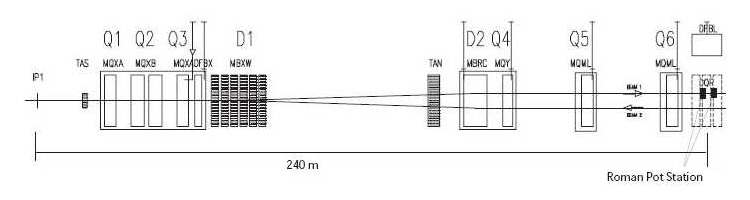}
\end{minipage}
\caption{The LHC magnetic lattice.}
\label{fig:lhc}
\end{figure*}

The LHC accelerates two proton beams. The one performing the clockwise rotation is called  \textit{beam1}, and the other one -- \textit{beam2}. The
beams traverse the LHC magnetic lattice in separate beam pipes which merge into a single one about 140~m away from the IP. A schematic 
view of the LHC lattice in the vicinity of the ATLAS IP is shown in Fig.~\ref{fig:lhc}.
The quadrupole magnets are labelled with ``Q'', and the dipole ones -- with ``D''. The Q1--Q3 quadrupoles serve as the final focusing and emittance 
matching triplet, while the D1 and D2 dipoles are used to separate the proton beams. In the analysis presented below the LHC magnetic lattice plays a role 
of a spectrometer for the scattered forward particles.  To protect the magnets against the radiation coming from the Interaction Point the Target Absorber 
Secondaries (TAS) and the Target Absorber Neutral (TAN) were installed in front of the final focusing triplet and the D2 magnet, respectively. In addition, 
there are several corrector magnets, beam position monitors and collimators, not shown in the Fig.~\ref{fig:lhc}.

\subsection{Zero Degree Calorimeters}

Zero Degree Calorimeter (ZDC) \cite{zdc} is a compact detector foreseen to measure neutral particles produced at very large pseudorapidities. ATLAS 
uses two such devices located symmetrically on both sides of the IP at the distance of 140~m in the slit of the TAN absorber. The ZDC is segmented 
longitudinally with the first segment dedicated to the measurement of electromagnetic radiation. The energy resolution of the calorimeter for photons is 
$\sigma/E = 0.58/\sqrt{E}+0.02$ and for neutrons $\sigma/E = 2.1/\sqrt{E}+0.12$ with the particle energy measured in~GeV. The calorimeter delivers 
also information on the position of a particle impact point. Throughout this work, it was assumed that the ZDC spatial resolution is 1~mm for photons and 
3~mm for neutrons, irrespectively of the incident particle energy.  These values are the upper bounds of the spatial resolutions estimated during the 
ZDC tests -- see \cite{zdc} for details. The active area of the ZDC face as seen from the ATLAS experiment IP  is  of rectangular shape with dimensions 
94~mm by 88~mm in horizontal and vertical directions, respectively.

\subsection{Absolute Luminosity for ATLAS -- ALFA}
\label{sec:alfa}
The Absolute Luminosity for ATLAS (ALFA) system \cite{alfa} was designed to measure elastically scattered protons. Such protons are scattered at very 
small polar angles and escape from the main ATLAS detector volume through the accelerator beam pipe and traverse the magnetic lattice of the machine. 
They can be registered in dedicated devices only. Typically, such detectors use the roman pot technology which allows the precise positioning of the 
detectors inside the beam pipe and in consequence a precise measurement of small scattering angles. 

It is quite clear that the acceptance of the considered detectors strongly depends on both the properties of the magnetic spectrometer, here the LHC, and the position 
of the detector with respect to the beam. The latter is quantified by a crucial parameter -- the distance between the detector active edge and the beam. 
This parameter determines minimal value of the measurable scattering angle. 

To minimise the influence of the beam angular divergence at the IP on the measurement one selects the large $\beta^\star$ optics. 
Therefore, the ALFA detectors are foreseen to work during the dedicated periods of the LHC running. 

There are four ALFA stations placed symmetrically w.r.t. the IP at the distances of 237~m and 251 m. Each station consists of two roman pots which 
enable an independent vertical insertion of the detector active parts into the accelerator beam pipe. Each roman pot contains one main and two overlap 
detectors. The main detector is made of 20 layers of scintillating fibers arranged in U-V geometry and contains also the trigger counter consisting of two 
layers of a scintillator. The spatial resolution of the scattered proton trajectory measurement in the main detector is about 30 $\mu$m. The overlap 
detectors consist of horizontal fibers are are dedicated to the measurement of the distance between the upper and bottom main detectors. In the 
following, the distance between the edge of the active part of the main detector and the outer wall (beam side) of the roman pot was assumed to be 
0.5~mm. The energy of the scattered proton registered by the ALFA stations can be determined with resolution of about 40~GeV \cite{maciek}. 

In the case of the 90~m optics, the ALFA detectors acceptance is above 40\% for the scattered protons having the relative energy loss, $\xi$  within the 
range $(0.0, 0.2)$ and with the transverse momentum, $p_T < 1.0$ ~GeV/c. However, it reaches above 80\% for the protons with 
$p_T \approx 0.5$~GeV/c (see \cite{maciek} for more details). One should recall that the lower cut on the proton transverse momentum depends on the 
distance between the detector and the beam.

\section{Properties of Final State}

The momentum conservation in the bremsstrahlung process and the softness of the exchange give: 
\begin{equation}
E_{2} \approx E_\gamma+E_1,
\label{eq:econs}
\end{equation}
and 
\begin{equation}
\vec{p_2} + \vec{p_\gamma}+\vec{p_1} = \vec0,
\label{eq:pcons}
\end{equation}
where $E_1 $ ($\vec{p_1}$)  is the radiating proton energy (momentum) after the emission, $E_\gamma$ ($\vec{p_\gamma}$) is the photon energy 
(momentum) and $E_2 $ ($\vec{p_2}$)  is the energy (momentum) of the proton scattered into the hemisphere opposite to the one containing the emitted 
photon. Since the radiation occurs in the elastic $pp$ scattering then  $E_1 = E_{beam} $ ($\vec{p_1} = \vec{p_{beam}}$). In addition, similarly to the 
electromagnetic bremsstrahlung, the final state photon angular distribution is driven by the colliding proton Lorentz factor $\gamma = E_{beam}/m_p$ 
$$ \frac{d\sigma}{d\Theta_\gamma} \sim \frac{\Theta_\gamma} {\left(\frac{1}{\gamma^2}+\Theta_\gamma^2\right)^2},$$
 \noindent
where $\Theta_\gamma$ is the polar angle of the emitted photon and $m_p$ is the proton mass \cite{jackson}. 

The average value of the photon polar angle is proportional to $1/\gamma$. In the case of the LHC running at the centre of mass energy 
$\sqrt{s} = 13000$~GeV $1/\gamma \approx 144$~$\mu$rad and corresponds to the photon pseudorapidity\footnote{The pseudorapidity of a particle is 
defined as $\eta = -\ln{\tan{\Theta/2}}$, where $\Theta$ is the polar angle of a particle.}, $\eta \approx 9.5$. In addition, the scattered proton angular
distribution is very narrow due to the large value of the nuclear slope parameter at high energies and its average value is $\sim\!1/(4\gamma)$  at the 
LHC energies.\\
The process cross-section is of the order of a few microbarns at the LHC energies.

\begin{figure}[h]
\begin{minipage}{\columnwidth}
\centering
\includegraphics[width=0.8\columnwidth]{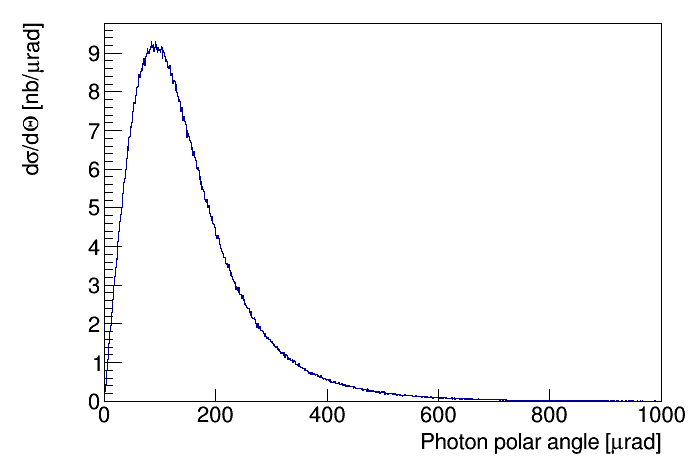}
\end{minipage}
\caption{The bremsstrahlung photon polar angle distribution.}
\label{fig:gtheta}
\end{figure}
\begin{figure}[h]
\begin{minipage}{\columnwidth}
\centering
\includegraphics[width=0.45\columnwidth]{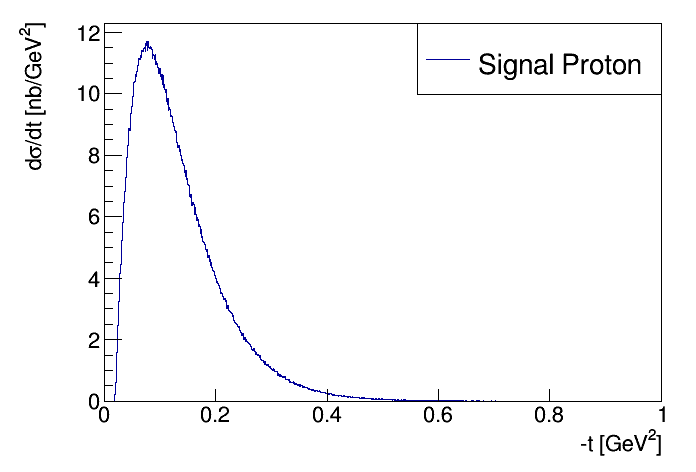}
\includegraphics[width=0.45\columnwidth]{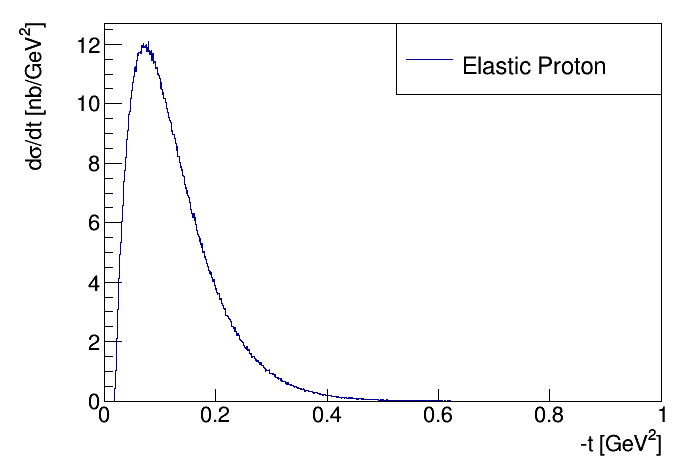}
\end{minipage}
\caption{The $t$  distribution of the radiating proton.}
\label{fig:tdist}
\end{figure}

Figure \ref{fig:gtheta} shows the distribution of the polar angle of the bremsstrahlung photon. The mean value of the photon emission angle is close to 
150~$\mu$rad. In Fig.~\ref{fig:tdist} the distributions of the four-momentum transfer squared, $ t$, of  the radiating (or ``signal'') proton (left) and that 
of the elastic one (right) are shown. These distributions well match the $t$-range accepted by the ALFA detectors~\cite{atlas, atlasel}. The photon energy 
spectrum (not shown) is of a typical $\sim\!1/E_\gamma$ shape and that of the proton relative energy loss is proportional to $1/\xi$. The final state 
particles are emitted in the forward directions. For the LHC running with $\beta^\star = 90$~m both final state protons can be registered with the help of 
the ALFA stations. Therefore, the signal signature is defined as the three-particle final state with the photon registered in the ZDC and the scattered 
protons in the ALFA stations. One should note that the presently assumed signal signature differs significantly from that discussed in \cite{app2} where the 
diffractive bremsstrahlung observation was based on the registration of a photon in the ZDC and a single proton in the AFP detectors located in the same 
hemisphere of the reaction.

\section{Analysis}

\subsection{Signal}
\label{sec:signal}

A sample of diffractive bremsstrahlung events was generated using the Monte Carlo generator based on GenEx \cite{kycia1} which was devised to 
describe the production of the low multiplicity exclusive states. The GenEx extension implements the calculations of \cite{szczurek}. A sample of three 
milion diffractive bremsstrahlung events with the photon energy between 100~GeV and 1500~GeV was generated. The value of the cross-section 
reported by the generator within this kinematic range is $1.7514\pm0.0006~\mu$b.

The final state particles were transported through the LHC lattice. The transport of the protons was performed using parameterisation \cite{rafal, maciek} 
of the MAD-X~\cite{mad} results. The differences between the MAD-X and the parameterisation results are much smaller than the spatial resolutions of 
the experimental apparatus. In the calculations the beam and interaction vertex properties were taken into account. One should note that the beam 
chamber geometry is of rather small importance owing to the low values of the transverse momenta of the final state particles. The ZDC response to the 
incident photon was calculated using the resolution mentioned above. Also, the photon impact position was smeared according to the reported spatial 
resolution of the ZDC calorimeters. Similarly, the positions of the scattered protons as measured by the ALFA detectors were smeared with appropriate 
resolutions. 

One should note that the measurement proposed below belongs to the class of exclusive measurements so speaking in general terms, the signal event 
was defined by the following requirements:
\begin{itemize}
\item energy of a photon reconstructed in the ZDC, $E_{\gamma,ZDC}$, within the range 130~GeV to 1500~GeV,
\item presence of a proton in both hemispheres. Each proton it was requested to be registered in both ALFA stations belonging to a given hemisphere.
\end{itemize} 
 
The above conditions suggest possible configuration of  the trigger. Namely, a double side (two tracks) ALFA trigger enhanced with the requirement of 
an electromagnetic energy deposit within the ZDC corresponding to at least 100 GeV photon energy. One should note that the rate of such a trigger would 
not be greater than the rate of the elastic ALFA trigger. In case of the low luminosity running associated with high-$\beta^\star$ running it will not 
challenge the bandwidth of the existing trigger and data acquisition systems.  

\subsection{Simulation of Background Processes}

The background processes were generated using \textsc{Pythia~8} Monte Carlo \cite{pythia}. The generated sample includes single and double diffractive 
dissociation\footnote{It was checked with Pythia that the non-diffractive processes contribution to the background after the final selection is negligible.} 
and corresponds to the cross-section of 21.4 mb. A sample of $10^9$ events was generated. These events underwent the simulation and reconstruction 
procedures mentioned in Sec. \ref{sec:signal}. 

As the source of the background all events producing the signal-like signature were considered \textit{i.e} events passing the criteria described in previous 
section. This means also that the central detector should not register any particle. Therefore, an event was rejected from the analysis if it satisfied at least 
one of the following requirements:
\begin{itemize}
\item presence of a charged particle with the transverse momentum $p_T > 1$~GeV and $|\eta| < 2.5$ -- the inner tracker veto;
\item presence of a particle with energy $E > 1$~GeV and  $|\eta| < 4.8$ -- the calorimeter veto,
\item presence of a neutral hadron with the energy reconstructed in any ZDC greater than 30~GeV,
\item the electromagnetic energy reconstructed in the ZDC belonging to the non-signal hemisphere larger than 30~GeV.
\end{itemize}
\noindent
The above criteria reflect the performance figures of the ATLAS detector. The first two selection cuts efficiently remove non-diffractive and high-mass 
diffractive processes, while the other two play an important role in the reduction of the double diffraction contribution.  

\begin{figure}[h]
\begin{minipage}{\columnwidth}
\centering
\makebox[0.45\columnwidth][l]{\  }\\
\includegraphics[width=0.55\columnwidth]{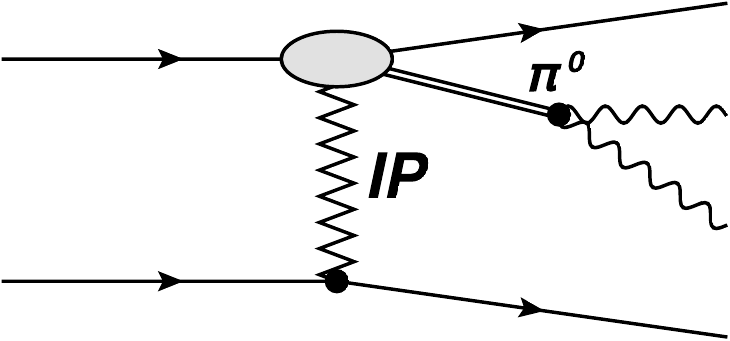}\\
\makebox[0.35\columnwidth][l]{\ }
\end{minipage}
\caption{The $\pi^0$-strahlung process diagram. The blob indicates a mixture of different mechanisms of the pion emission.}
\label{fig:pio-strahl}
\end{figure}

It was observed that the dominating source of the remaining background events is $\pi^0$-strahlung:
$$p+p\rightarrow p+ \pi^0 +p$$
shown in Figure~\ref{fig:pio-strahl}. The estimated cross-section for this process is about 300~$\mu$b~\cite{szczurek2}.

The events were required to contain a \textit{single} photon in the ZDC. For each photon within the ZDC acceptance its 
reconstructed position was calculated assuming the ZDC spatial resolution. The events containing two or more photons were removed from the analysis if 
the maximum distance between their impact positions was larger than 6 mm. One should note that the distance between the photons emerging from the 
$\pi^0$ decay at the ZDC face is not smaller than 5 mm in the $\pi^0$ energy range seen in the present analysis and which is a consequence of the 
$\pi^0$ mass.

\subsection{Optimisation of the Background Reduction}

The described above event selection for the signal and background samples leads to a very low value of the signal to background ratio -- well below 1.
This is also a consequence that the ALFA detectors are merely used as the YES/NO-taggers. Therefore, in the following, an improvement of this ratio is 
attempted. This was performed using properties of the signal and background final state particles. 

At first, the distributions of the photon polar angle were studied. The distributions of the photon impact position on the ZDC face obtained for the signal and 
background samples were compared. 
\begin{figure}[h]
\begin{minipage}{\columnwidth}
\centering
\includegraphics[width=\columnwidth]{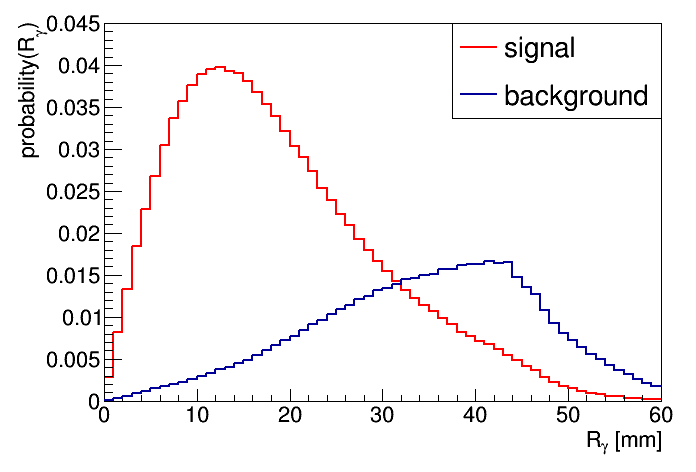}
\end{minipage}
\caption{The distribution of the distance, $R_\gamma$, between the reconstructed photon impact point at the ZDC and the beam projection (areawise 
normalised). The red line - the signal event sample, the blue line - the background sample (colour on-line).}
\label{fig:rsig}
\end{figure}
\noindent
In particular, the distributions of the distance, $R_\gamma$, measured in the ZDC face plane, between the photon impact point and the linearly 
extrapolated position of the beam were studied. These distributions are shown in Fig. \ref{fig:rsig} for events passing the above outlined selections.

As can be observed, these two distributions clearly differ. The signal one after an initial increase reaches a maximum located at about 14 mm and for 
larger values of $R_\gamma$ rapidly decreases.  In the case of the distribution constructed for the background events the initial increase is much slower 
and the maximum is observed for $R_\gamma \approx 40$~mm. The shape of the former distribution reflects the angular distribution of the emitted 
photons (confront also Fig.~\ref{fig:gtheta}) while the shape of the latter one follows from the much wider angular distribution of the neutral pions 
produced in $pp\rightarrow pp \pi^0$ reaction \cite{szczurek2} and is additionally smeared due to kinematic properties of the photons from the $\pi^0$ 
decay. The difference described above will serve as an efficient background discrimination tool.

\begin{figure}[h]
\begin{minipage}{\columnwidth}
\centering
\includegraphics[width=\columnwidth]{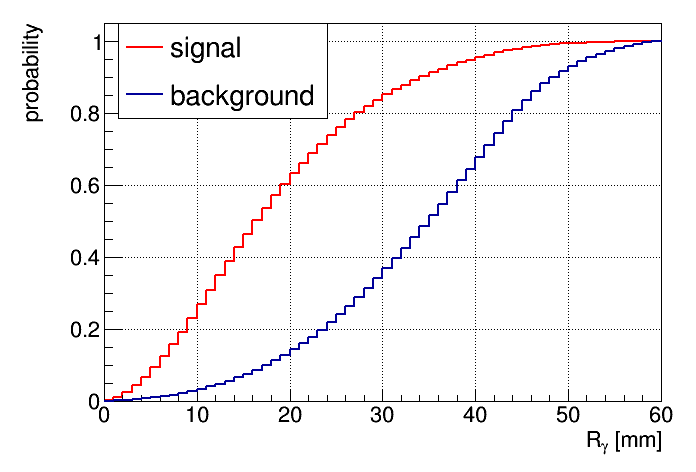}
\end{minipage}
\caption{The distribution of the probability that the distance, $R_\gamma$, between the reconstructed photon impact point at the ZDC and the beam 
projection  is smaller than a certain value. The red line - the signal event sample, the blue line - the background sample (colour on-line).}
\label{fig:risig}
\end{figure}

The distributions shown in Fig. \ref{fig:rsig} were used to construct the distributions of the probability that the discussed distance is smaller than a certain 
value and  the results of the calculations are compared in  Fig.~\ref{fig:risig}. 

Inspection of  Fig. \ref{fig:risig} shows that the requirement that $R_\gamma$ is smaller than 30~mm retains about 85\% percent of the signal events 
and rejects about 65\% of the background events. Imposing such a cut leads to the visible cross-sections of 1.16~$\mu$b and 12.8 $\mu$b for the signal 
and the background, respectively. This corresponds to the signal to background ratio of about 1/10. It is worth stressing that such a low value is obtained 
without any requirement on the event exclusivity, in particular without exploiting the four-momentum conservation. On the other hand one should observe 
that the use of the energy and momentum conservation allows a further optimisation of the background reduction which can be performed in the following 
way. This procedure makes use of the properties of the final state particles produced in diffractive bremsstrahlung process. Namely, the proton scattered 
into the hemisphere opposite to the ``signal'' one, to a very good approximation, preserves the energy of the beam \textit{i.e.} has properties resembling 
a proton scattered in elastic process. Therefore, its transport through the structures of the LHC can be computed using the machine optics (transport 
matrices) designed for the measurement of the elastic $pp$ scattering. This also means that the backward transport -- from the ALFA station to the ATLAS 
IP -- can be calculated in a simple manner. This allows a full determination of the elastic proton transverse momentum at the IP using its trajectory position 
measured by the ALFA detectors. In addition, reconstruction of the bremsstrahlung photon impact point on the ZDC face and the measurement of its 
energy delivers full information on the measured momentum of the photon. Assuming that there is a three-particle final state in the event and applying Eq. 
(\ref{eq:pcons}), the four-momentum vector of a pseudo-proton (pseudo-p$_1$ particle) can be constructed\footnote{Observe that this pseudo-particle is 
an analog of the proton scattered into the signal hemisphere}. One should anticipate that the energy and momentum distributions of such a 
pseudo-particle can, in principle, be different for the signal and background samples as a consequence of the presence in the background events of 
additional final state particles escaping the detector acceptance. In the next step its position at the ALFA stations is calculated using the transport 
parametrisation and next the distance, $R_p$, between its impact point and that of the signal proton is determined. It should be noted that in 
bremsstrahlung events the pseudo-p$_1$ particle four-momentum should be very close to that of the genuine signal proton. Possible difference is a result 
of various resolutions. On the other hand in the case of a background event the differences are expected to be large so are the $R_p$ values. 
This is a consequence of the lack of the momentum conservation for background events. One expects that the background events contain more than three
final state particles. These additional ones have to escape the detectors' acceptance (so the event passes the selection criteria) introducing the missing 
momentum in the event and hence large smearing of the pseudo-proton momentum and large $R_p$ values. One should note that this is so also true for
exclusive $pp\rightarrow pp+\pi^0$ events since the selection requires a single photon in the ZDC, \textit{i.e.} a part of the $\pi^0$ momentum is not 
measured. 

\begin{figure}[h]
\begin{minipage}{\columnwidth}
\centering
\includegraphics[width=\columnwidth]{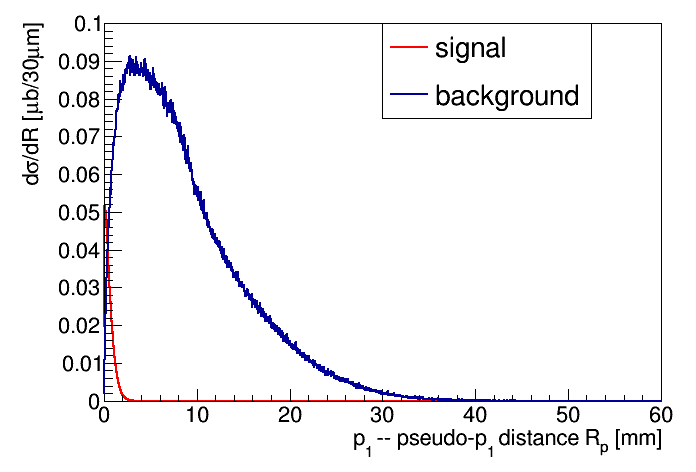}
\end{minipage}
\caption{The distribution of the probability that the distance, $R_p$, between the $p_1$ proton and pseudo-$p_1$ positions in the ALFA detector (see 
text). The red line - the signal event sample, the blue line - the background sample (colour on-line).}
\label{fig:prsig}
\end{figure}
\begin{figure}[hb]
\begin{minipage}{\columnwidth}
\centering
\includegraphics[width=\columnwidth]{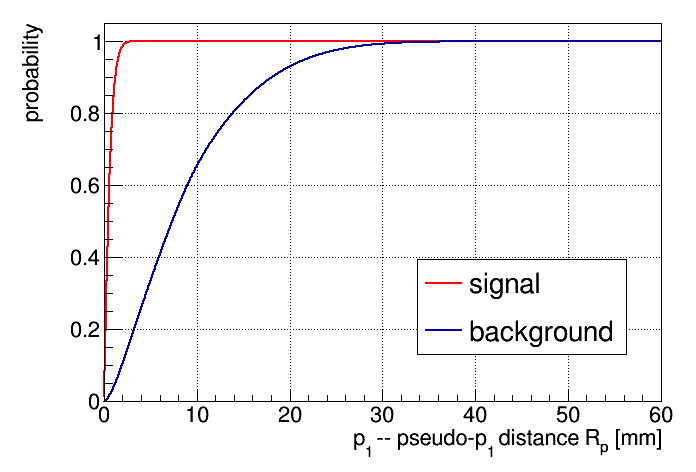}
\end{minipage}
\caption{The distribution of the probability that the distance, $R_p$, between the $p_1$ proton and pseudo-$p_1$ positions in the ALFA detector (see 
text) is smaller than a certain value. The red line - the signal event sample, the blue line - the background sample (colour on-line).}
\label{fig:prisig}
\end{figure}

The distributions of the $R_p$ distance calculated for the signal and the background samples are shown in Figure~\ref{fig:prsig}. As can be observed the 
signal distribution is almost totally confined within 2 millimetres. This is in contrast with the behaviour observed for the background sample distribution 
which after a steep rise reaches its maximum for the distance of about 4~mm and then relatively slowly decreases extending up to 30-40~mm. To 
evaluate the rejection power of the cut on $R_p$ both distributions were converted into the probability distribution that $R_p$ is smaller than a given 
value. A plot comparing both probability distributions is shown in Figure~\ref{fig:prisig}.

As was already mentioned, the acceptance of the ALFA detectors crucially depends on the distance between the beam and the active edge of the 
detector. Usually, this distance is measured in the units of the transverse size of the beam, $\sigma_{beam}$, at the detector location (see 
Table~\ref{tab:param}). In addition, a constant distance of 0.5~mm covering 0.3~mm of the roman pot floor thickness and 0.2~mm of the distance 
between the inner plane of the floor and the detector active edge. The total distance between the beam and the active edge of the detector is: 
$$d_{tot} = 0.5+n\cdot\sigma [mm].$$ 
In the calculations $n$ was varied between 10 and 30 in steps of 5. In addition, the geometrical limit, along the horizontal direction, of $|x| \leq 14$~mm
was taken into account. 

As mentioned above, the visible cross-section, and hence the signal to background ratio, is a function of the cut on $R_p$ and the value of $n$. 
Tables~\ref{tab:visxs2dim} and \ref{tab:sb2dim} present the signal visible cross-section and the signal to background ratio as a function of  
$R_p$ and $n$, respectively.
\begin{table}[h]
\centering
\caption{The signal visible cross-section in nanobarns as a function of $R_p$ and $d_{tot} = 0.5+n\cdot\sigma_{beam}$ for the LHC running with 
$\beta^\ast = 90$ m. }
\label{tab:visxs2dim}

\begin{tabular*}{\columnwidth}{@{\extracolsep{\fill}}l|rrrrr@{}}
\hline
$R_p$[mm]& \multicolumn{5}{c}{$n$ [$\sigma_{beam}$].  $d_{tot} = 0.5+n\cdot\sigma_{beam}$.}\\
\cline{2-6}
 &	10	&	15	&	20	&	25	&	30\\
\hline
0.1	&	50	&	44	&	39	&	34	&	30	\\
0.2	&	114	&	101	&	89	&	77	&	67	\\
0.3	&	174	&	155	&	136	&	119	&	104	\\
0.4	&	231	&	206	&	182	&	160	&	139	\\
0.5	&	281	&	252	&	224	&	197	&	171	\\
0.6	&	325	&	292	&	261	&	230	&	201	\\
0.9	&	424	&	385	&	347	&	309	&	271	\\
1.0	&	447	&	408	&	368	&	328	&	288	\\
1.2	&	482	&	441	&	400	&	357	&	315	\\
1.5	&	514	&	472	&	429	&	385	&	341	\\
1.8	&	530	&	488	&	445	&	400	&	354	\\
2.0	&	535	&	494	&	450	&	405	&	359	\\
\hline
\end{tabular*}
\end{table}

\begin{table}[h]
\centering
\caption{The signal to background ratio as a function of $R_p$ and $d_{tot} = 0.5+n\cdot\sigma_{beam}$ for the LHC running with 
$\beta^\ast = 90$~m. }
\label{tab:sb2dim}
\begin{tabular*}{\columnwidth}{@{\extracolsep{\fill}}l|rrrrr@{}}
\hline
$R_p$[mm]& \multicolumn{5}{c}{$n$ [$\sigma_{beam}$]. $d_{tot} = 0.5+n\cdot\sigma_{beam}$.}\\
\cline{2-6}
 &	10	&	15	&	20	&	25	&	30\\
\hline
0.1	&	14.7	&	14.8	&	15.3	&	15.3	&	15.3	\\
0.2	&	8.6	&	8.5	&	8.4	&	8.4	&	8.5	\\
0.3	&	6.4	&	6.3	&	6.2	&	6.2	&	6.3	\\
0.4	&	4.9	&	4.9	&	4.9	&	4.9	&	4.9	\\
0.5	&	4.1	&	4.1	&	4.1	&	4.1	&	4.1	\\
0.6	&	3.5	&	3.6	&	3.6	&	3.6	&	3.7	\\
0.9	&	2.4	&	2.5	&	2.5	&	2.6	&	2.6	\\
1.0	&	2.2	&	2.2	&	2.3	&	2.3	&	2.4	\\
1.2	&	1.8	&	1.8	&	1.9	&	1.9	&	2.0	\\
1.5	&	1.3	&	1.4	&	1.4	&	1.5	&	1.5	\\
1.8	&	1.0	&	1.1	&	1.1	&	1.2	&	1.2	\\
2.0	&	0.9	&	0.9	&	1.0	&	1.0	&	1.0	\\
\hline
\end{tabular*}
\end{table}

The signal visible cross-section decreases with increasing $d_{tot}$ for a fixed value of $R_p$. It varies between tens of nanobarns for $R_p = 0.1$~mm 
and few hundred nanobarns for $R_p = 2.0$~mm. For a given value of $n$ ($d_{tot}$) the signal visible cross-section naturally increases with increasing 
$R_p$ and eventually saturates.
 
\begin{figure}[h]
\begin{minipage}{\columnwidth}
\centering
\includegraphics[width=\columnwidth]{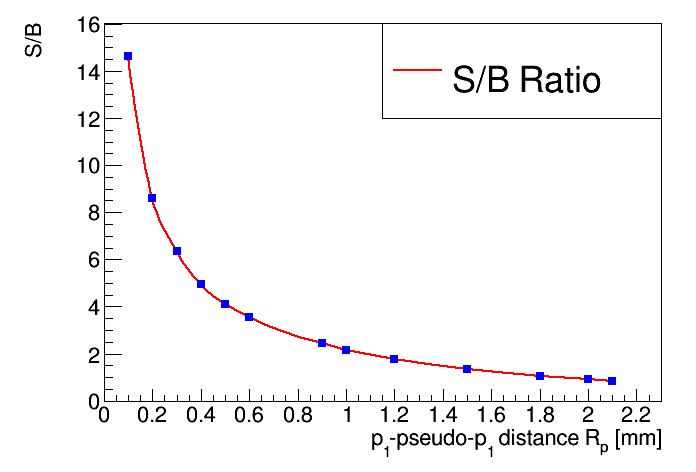}
\end{minipage}
\caption{The signal to background ratio  for $d_{tot} = 0.5+10\cdot\sigma$ and $R_p = 0.5$~mm for the LHC working with $\beta^\ast = 90$ m.
}
\label{fig:sb-rp}
\end{figure}

Table \ref{tab:sb2dim} shows that the signal to background ratio very weakly depends on $n$ ($d_{tot}$) for a fixed value  of $R_p$. However, for a given
value of $n$ it decreases from about 15 to about 1 with $R_p$ increasing  from 0.1~mm to 2.0~mm.  This is illustrated in Fig.~\ref{fig:sb-rp} 
which shows the  signal to background ratio calculated for $d_{tot} = 0.5+10\cdot\sigma_{beam}$ as a function of $R_p$. It is worth noting that the 
signal visible cross-section ranges between about 3\% and 30\% of the diffractive bremsstrahlung cross-section within the discussed span of $R_p$.

\begin{figure}[h]
\begin{minipage}{\columnwidth}
\centering
\includegraphics[width=\columnwidth]{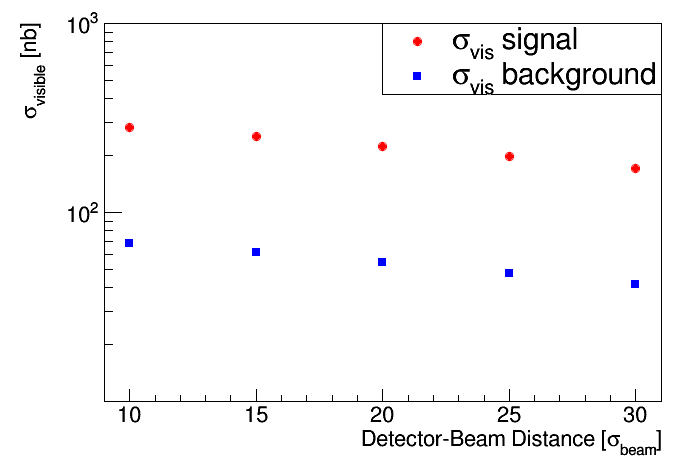}
\end{minipage}
\caption{The visible cross-sections for the signal and background as a function of $n$  for $R_p = 0.5$~mm and $\beta^\ast = 90$ m. 
}
\label{fig:selvisxs}
\end{figure}

Table \ref{tab:visxs} lists the signal and background visible cross-sections as a function of the detector--beam distance for the signal and background 
samples for $R_p = 0.5$~mm. Also, the signal to background ratio is given in the table and the visible cross-sections are depicted in 
Figure~\ref{fig:selvisxs}.

\begin{table}[h]
\centering
\caption{The visible cross-sections in nb for the signal and background for different distances of the detector active part and the beam for 
$R_p = 0.5$~mm and $\beta^\ast = 90$ m. }
\label{tab:visxs}
\begin{tabular*}{\columnwidth}{@{\extracolsep{\fill}}lrrrrcc@{}}
 &	{Signal}  & {Background}&{S/B ratio}\\
distance [mm]& {cross-section [nb] } & {cross-section [nb]}  &  \\
\hline
$0.5+10\cdot\sigma_{beam}$ &281& 68&  4.12 \\
$0.5+15\cdot\sigma_{beam}$ &252& 61 & 4.10\\
$0.5+20\cdot\sigma_{beam}$ &224& 54 & 4.11\\
$0.5+25\cdot\sigma_{beam}$ &197& 48 & 4.14\\
$0.5+30\cdot\sigma_{beam}$ &171& 41 & 4.13 \\
\hline
\end{tabular*}
\end{table}

As can be seen from Figure \ref{fig:selvisxs}  the visible cross-sections decrease by about 1/3 with increasing detector-beam distance (confront also 
Table~\ref{tab:visxs2dim}). However, their dependence on $d_{tot}$ is very similar resulting, in fact, a practically constant the signal to background ratio. 

\subsection{Results}

The signal visible cross-section values range between few tens and few hundred of nanobarns while the background one varies between few and few 
hundred nanobarns. The visible cross-sections  are decreasing functions of the detector-beam distance. However, they increase with increasing value of 
the cut on the distance ($R_p$) between the signal proton position and the pseudo-signal proton position in the ALFA station. The resulting signal to 
background ratio very weakly depends  on the detector-beam distance, however, it decreases with increasing value of $R_p$.

One should note that these cross-section values imply the signal (background) rates varying between about 0.3~Hz (0.02~Hz) and 50~Hz (50~Hz) for the
instantaneous luminosity range of $10^{30}$~cm$^{-2}$s$^{-1}$ to $10^{32}$~cm$^{-2}$s$^{-1}$. The rates for given values of $R_p$ and $n$ 
can be easily estimated using Tables~\ref{tab:visxs2dim} and~\ref{tab:sb2dim}. 

It has to be stressed that the above discussed values were obtained with an implicit assumption of a single interaction per bunch crossing and assuming
that the luminosity is evenly distributed among all bunches. These conditions can be and in fact are achieved during the LHC runs dedicated to the 
measurements using the ALFA stations. Nevertheless, there exist a finite probability of the multiple $pp$ interaction during a single bunch-crossing. 
This probability increases with increasing value of the instantaneous luminosity of the accelerator. To take it into account one has to multiply the visible 
cross-section by correction factor which is practically 1 for luminosities below $10^{30}$ cm$^{-2}$s$^{-1}$, reaches the value of about 0.9 for 
luminosity of 
$10^{31}$ cm$^{-2}$s$^{-1}$, the value of about 0.4 for $10^{32}$ cm$^{-2}$s$^{-1}$ and for larger instantaneous luminosities rapidly decreases.\\
\noindent
The signal visible cross-section implies the collection of the diffractive bremsstrahlung samples of  few thousands to several thousands events during
 a single LHC fill lasting typically about 10 hours.

\section{Summary and Conclusions}

Feasibility studies of the diffractive bremsstrahlung measurement at the centre of mass energy of 13 TeV at the LHC running with $\beta^\ast = 90$~m
were performed. The postulated measurement is based on the registration of the bremsstrahlung photon in the ZDC and both protons in the ALFA stations. 
The expected visible cross-section depends on the distance between the detector and the beam and the cut on the distance ($R_p$) between the positions 
of the signal and pseudo-signal protons in the ALFA stations. The visible cross-section decreases with increasing distance between the detector and the 
beam and increases with increasing value of $R_p$. The expected rates allow the collection of large event samples during a single LHC fill (about 10 
hours) with low and moderate instantaneous luminosity. The visible cross-section for the background shows similar dependence on the discussed 
distances.The resulting signal to background ratio decreases between about 15 and 1 with $R_p$ value increasing from 0.1~mm to 2.1~mm and very 
weakly depends on the detector-beam distance.

One should note that the machine conditions, for example a coherent background like beam halo, will influence potential measurement. Therefore, 
further optimisation of the analysis cuts has to take into account actual running conditions during the data taking period.  

Qualitatively, these studies apply also to CMS/TOTEM detectors. However, one has to remember that the applied cuts and the values of the signal to 
background ratios depend on the actually used experimental set-up. 

\section*{Acknowledgements}
\ \\
This work was supported in part by Polish National Science Centre grant UMO-2015/18/M/ST2/00098.

\end{document}